\documentclass[%
reprint,
amsmath,amssymb,
aps,
]{revtex4-2}

\usepackage{graphicx}
\usepackage{dcolumn}
\usepackage{bm}

\usepackage[utf8]{inputenc}
\usepackage[T1]{fontenc}
\usepackage{booktabs, array, mathptmx, float, tabularx, booktabs, lipsum, amsmath,multirow}
\usepackage{siunitx, xcolor}
\usepackage[version=4]{mhchem}
\graphicspath{{figs/}{figsgaoerb/}} 
\usepackage[colorlinks,linkcolor=blue,anchorcolor=blue,citecolor=blue]{hyperref}

\begin{document}


\title{Squeezed Displaced Schr\"{o}dinger-cat state as a signature of the $\mathcal{P}\mathcal{T}$-symmetry phase transition}


\author{Yuetao Chen}
\author{Shoukang Chang} 
\author{Shaoyan Gao}
\email{gaosy@xjtu.edu.cn}
\affiliation{MOE Key Laboratory for Nonequilibrium Synthesis and
	Modulation of Condensed Matter, Shaanxi Province Key Laboratory of Quantum
	Information and Quantum Optoelectronic Devices, School of Physics, Xi'an
	Jiaotong University, 710049, People's Republic of China.}




\begin{abstract}
Parity-time ($\mathcal{P}\mathcal{T}$) symmetric systems are gain-loss systems whose dynamics are governed by non-Hermitian Hamiltonians with degeneracies at exceptional-points (EPs) and has been studied in various photonic, electrical, mechanical systems, and so on. However, it is still an open question how to capture $\mathcal{P}\mathcal{T}$ symmetry phase transition in electronic system where the transport properties of electron will be dramatically effected. Fortunately, the hybridization between photon and electron offers a novel way not only to control but also probe material properties. Here, we investigate a cavity coupled to a non-Hermitian Su-Schrieffer-Heeger (SSH) chain within mean-field ansatzs. We find that Squeezed Displaced Schrodinger cat (SDSc) will emerge with high fidelity in cavity ground state when $\mathcal{PT}$-symmetry is broken and the fidelity will experience a sharp drop from almost 1 to 0 as $\mathcal{PT}$ symmetry recovers. Additionally, in semiclassical limit, we find that there exists local extrema at two sides of x=0 in semiclassical photon Hamiltonian $H_{\mathrm{eff}}(x,p)$, a clear signature of the emergence of SDSc state in cavity ground state. Thus, the appearance of SDSc state can be used to capture $\mathcal{PT}$-symmetry phase transition which can not be modified by cavity mode. Besides, we exploit the cavity ground state to estimate the phase in the optical interferometer, and show that the quantum Fisher information and nonclassicalty will sharply decline at EPs. This reveals that $\mathcal{PT}$-symmetry breaking in electronic materials can also be captured by the quantum Fisher information and nonclassicalty in phase estimation.  
\end{abstract}
\maketitle


\section{Introduction}
The $\mathcal{PT}$-symmetry phase transition, as a subset of non-Hermiticity which represents a fundamental departure from conventional quantum mechanics [1–10], manifesting a transition from a real energy spectrum to a complex one. Intrinsic $\mathcal{PT}$-symmetry breaking spontaneously emerges at EPs where nonreciprocal hopping strength is equal to reciprocal one in a sublattices, and eigenstates and eigenvalues coincide [11–19]. 
The presence of EPs gives rise to rich physical phenomena, including enhancing precision in quantum sensors [20], topological phase transitions [21–25], nonadiabatic transitions [26,27], unidirectional light propagation [28], and energy transport at macroscopic [29], among others. 
In recent years, significant attention has been drawn to realize $\mathcal{PT}$-symmetry phase transition, with demonstrations in optical system [30–32], electrical system [33], and mechanical setting [34]. However, apart from the interest of realizing $\mathcal{PT}$-symmetry phase transition systems in various fields, probing $\mathcal{PT}$-symmetry breaking is still an open question, particularly in electronic systems, where such breaking can exert a significant influence on electron transport. 

Fortunately, quantum Floquet presents a novel way for probing, manipulating, and fine-tuning material properties, leading for instance to the emergence of polaritons—hybrid light-matter excitations—exhibiting non-trivial topological characteristics [35–41], or to an anomalous Hall response in the presence of a circularly polarized field [42]. As the quantum version of classical Floquet engineering, quantum Floquet has been proposed to control materials
through quantum light without detrimental heating [43]. The fundamental idea involves embedding a material within an optical cavity and amplify the light-matter interaction [44, 45] due to the inverse square-root relationship between the coupling strength and the effective mode volume [45, 46]. This amplification can be achieved, for instance, through near-field enhancement effects [47]. Through this enhancement of the coupling, one can achieve the aim of classical Floquet engineering with only quantum vacuum fluctuations or a few photons in cavity. To meet the theoretical requirements, ultra-strongly coupled light-matter systems have been achieved through various implementation schemes, beginning with initial findings utilizing microwave and optical cavities [48, 49] where the properties of electronic materials can be well manipulated. Apart from controlling materials, the interaction between cavity photons and material electron also has an effect on cavity ground state, leading for example to the emergence of squeezed vacuum state [50] and displaced squeezed vacuum state [51] where cavity Hamiltonian can be renormalised by hopping parameter in hermitian electronic materials within mean-field ansatzs. Naturally, a basic question to be asked here is what is the fate of the cavity ground state coupling to electron in the presence of $\mathcal{PT}$-symmetry phase transition in non-Hermitian electronic materials and whether one can capture $\mathcal{PT}$-symmetry phase transition sensitively by a probe of cavity ground state. 

In this paper, we address this fundamental question by investigating the interplay of electron and quantum light by considering the non-Hermitian electronic SSH chain coupled to a single mode cavity through the full gauge
invariant Peierls phase. Electronic energy spectrum and cavity ground state can be obtained by diagonalizing electronic Hamiltonian and photon Hamiltonian within mean-field ansatzs where there is no correlation between electron and photon. From view of electronic degree, we shows that quantum fluctuations of the light field have no effect on $\mathcal{PT}$-symmetry phase transition but exhibit dramatic effects on the topological properties of the system which can be probed by photonic spectrum. However, from the perspective of photon degree, the results show that Squeezed Displaced Schr\"{o}dinger-cat (SDSc) state with high fieldity emerges as $\mathcal{PT}$-symmetry broken and then disappears immediately when $\mathcal{PT}$ symmetry recovers. This reveals that the emergence of SDSc state can be regarded as a signature of $\mathcal{PT}$-symmetry phase transition in electronic materials and is possible due to 
the appearance of local extrema at two sides of x=0 in semiclassical photon Hamiltonian $H_{\mathrm{eff}}(x,p)$.  

With the cavity ground state in our scheme, we also estimate the linear phase in the optical interferometer. We find that the quantum Fisher information (QFI) about the phase and the nonclassicality of cavity ground will experience dramatic drop at EPs. Thus, QFI and the nonclassicality of cavity ground can be also regarded as signatures of $\mathcal{PT}$-symmetry phase transition. 
 
This paper is organized as follows. In \ref{Hamiltonian}, we introduce
we describe the Hamiltonian of the light-matter coupled system. In \ref{spectrum}, we show the effect of quantum vacuum fluctuation in cavity on electronic spectrum in open boundary condition. In \ref{generation}, we  discuss about the generation of SDSc state as $\mathcal{PT}$-symmetry breaking and investigate the influence of $\mathcal{PT}$-symmetry phase transition on the fidelity between exact SDSc state and cavity ground state. Semiclassical photon Hamiltonian $H_{\mathrm{eff}}(x,p)$ will be studied for explaining the physics behind the emergence of SDSc state. In \ref{Phase}, we exploit the cavity ground state to estimate the phase of the optical interferometer, and study the QFI and nonclassicality of cavity ground state at two sides of EPs. Finally, \ref{Conclusion} presents the discussion and summary of the main findings in this work.

\section{Hamiltonian}
\label{Hamiltonian}
\textbf{Model.} We consider a 1D tight-binding non-Hermitian SSH\ model
coupled to a single mode cavity which is pictorially shown in Fig. 1. 
\begin{figure}[h]
	\label{Fig.1} \centering\includegraphics[width=0.95\columnwidth]{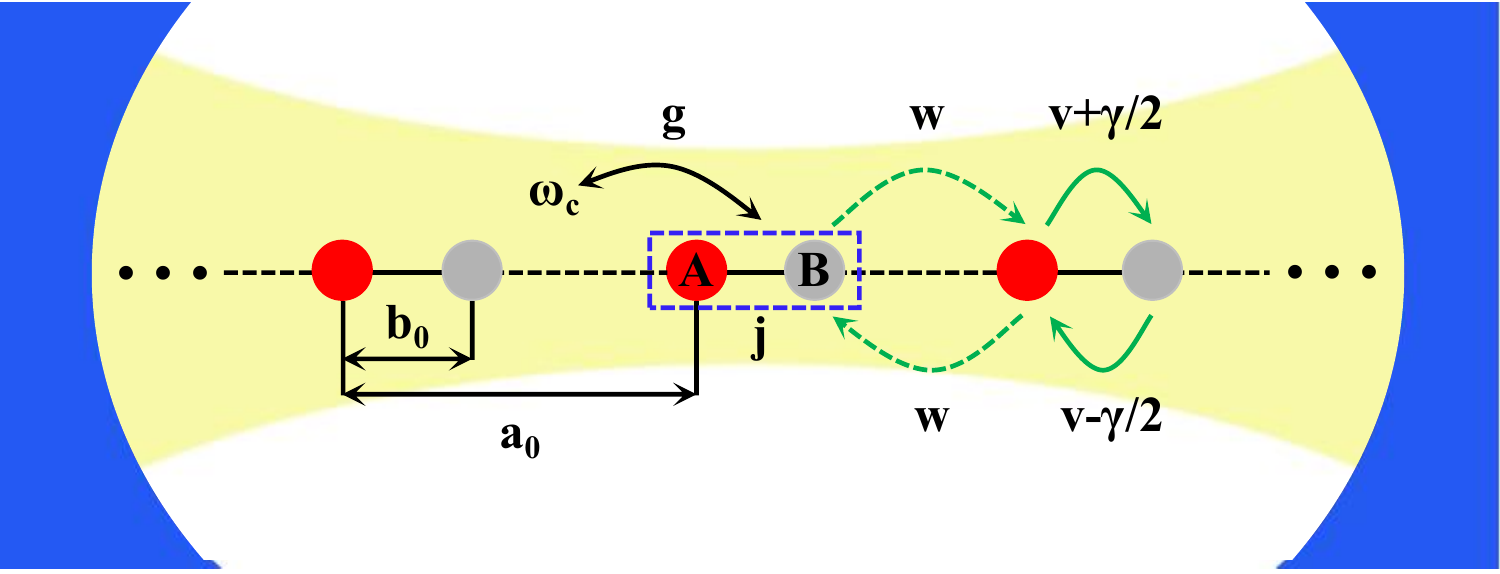}
	\caption{\textbf{Scheme of the non-Hermitian Su-Schrieffer-Heeger chain
			coupled to cavity.} A one-dimensional dimerized Su-Schrieffer-Heeger chain
		with nonreciprocal\ intracell $v\pm \protect\gamma /2$ (green solid line)
		and intercell $w$ (green dashed line) hopping amplitudes coupled to a single
		mode cavity with frequency $\omega _{c}$. The strength of the light-matter
		coupling is denoted by $g_{coupling}$. The lattice constant is given by $a_{0}$, and
		the distance between the sublattices A and B within the same unit cell
		(depicted by the blue dashed square) is $b_{0}$. }
\end{figure}
The above model describes spin-free electron nonreciprocal nearest-neighbor\
hopping in a one-dimensional chain of L units with two sublattices A and B. 
The corresponding Hamiltonian is
\begin{equation}
		\begin{split}
			H_{non-Hermitian-SSH} =(v+\frac{\gamma }{2})\sum_{j=1}^{L}c_{j,A}^{\dagger
			}c_{j,B}+(v-\frac{\gamma }{2})\sum_{j=1}^{L}c_{j,B}^{\dagger }c_{j,A}
			\\-w\sum_{j=1}^{L-1}c_{j+1,A}^{\dagger
			}c_{j,B}-w\sum_{j=1}^{L-1}c_{j,B}^{\dagger }c_{j+1,A}, \label{1}
		\end{split}
\end{equation}%
where $v(w)$ are lattice intracell (intercell) hopping strength and $\gamma $
is the nonreciprocal\ intracell hopping strength, meanwhile $c_{j,\zeta
}^{\dagger }(c_{j,\zeta })$ are the fermionic creation (annihilation)
operators at site $j$ and sublattice $\zeta =A,B$. $a_{0}$ and $b_{0}$ are
the lattice constant and intracell distance between $A$ and $B$ sites
correspondingly as shown in Fig. 1. The non-Hermitian SSH model displays
topological phase transitions $(v=\pm \sqrt{w^{2}+(\frac{\gamma }{2})^{2}})$
and $\mathcal{PT}$-symmetry breaking phase transitions $(v=\pm \frac{\gamma }{2})$ in
the absence of any light-matter interaction.\ It's worth noting that the two
phase transitions mentioned above are independent of the intracell distance $%
b_{0}$ without cavity. The Hamiltonian of the single mode cavity which is
coupled to the non-Hermitian SSH\ model reads\

\begin{equation}
	H_{phot}=\omega _{c}(a^{\dagger }a+\frac{1}{2}),  \label{2}
\end{equation}%
where $\omega _{c}$ is the mode frequency and the single cavity mode
annihilation (creation) operators are represented by $a(a^{\dagger })$
satisfying the Bose-Einstein commutation relation $\left[ a,a^{\dagger }%
\right] =1$. We couple the 1D non-Hermitian SSH model to the single cavity
mode by applying a unitary transformation $U$ to electronic Hamiltonian[47],
i.e.

\begin{equation}
	H=H_{phot}+U^{\dagger }H_{non-Hermitian-SSH}U,\   \label{3}
\end{equation}%
where the unitary transformation $U$ is defined as

\begin{equation}
	\ U=\exp (ieA\sum_{j,\zeta }r_{j,\zeta }c_{j,\zeta }^{\dagger }c_{j,\zeta }),
	\label{4}
\end{equation}%
where $r_{j,\zeta }$ represents the position of the sublattice, with $%
r_{j,A}=ja_{0}$ and $r_{j,B}=ja_{0}+b_{0}$. The full Hamiltonian under this
transformation can be written as\ (setting $a_{0}=1$)
\begin{equation}
	\begin{split}
		& H=(\nu+\frac{\gamma}{2})e^{i\frac{g_{coupling}}{\sqrt{L}}b_0(a+a^{\dagger})}\sum_{j=1}^{L}c_{j,A}^{\dagger}c_{j,B} \\
		& +(\nu-\frac{\gamma}{2})e^{-i\frac{g_{coupling}}{\sqrt{L}}b_0(a+a^{\dagger})}\sum_{j=1}^{L}c_{j,B}^{\dagger}c_{j,A} \\
		& -we^{i\frac{g_{coupling}}{\sqrt{L}}(1-b_0)(a+a^{\dagger})}\sum_{i=1}^{L-1}c_{j+1,A}^{\dagger}c_{j,B} \\
		& -we^{-i\frac{g_{coupling}}{\sqrt{L}}(1-b_0)(a+a^{\dagger})}\sum_{i=1}^{L-1}c_{j,B}^{\dagger}c_{j+1,A} \\
		& +\omega_{c}(a^{\dagger}a+\frac{1}{2}), \label{5}
	\end{split}
\end{equation}
where $A=\frac{g_{coupling}}{\sqrt{L}}(a+a^{\dagger })$ are related to the quantized
electromagnetic vector potential with the convention $e=\hbar =c=1$. It's
worth noting that the full light-matter interaction Hamiltonian obtained by
applying unitary transformation is equivalent to the Hamiltonian  under Peierls substitution[] where the hoppings amplitudes in Eq. (%
\ref{1})\ are dressed as $v\rightarrow v\exp (ieAb_{0}),\gamma \rightarrow
\gamma \exp (ieAb_{0})$ and $w\rightarrow w\exp \left[ -ieA(a_{0}-b_{0})%
\right] $. This phenomenon will play a key role in modifying the electron energy spectrum under open boundary conditions. Furthermore, the presence of non-Hermitian SSH chain will exert a discernible impact on the ground state of the cavity depending on the hopping parameters in non-Hermitian SSH chain. Strikingly,
we will show that the presence of nonreciprocal hopping in non-hermitian SSH
model can lead to the emergency of SDSc states with high
fidelity which can be regarded as a signature of the
$\mathcal{PT}$-symmetry breaking phase transition in non-Hermitian SSH chain and have great potential for applications in quantum information
processing and quantum metrology.

\section{The open-boundary energy spectrum of non-Hermitian SSH model in
	the presence of cavity}
\label{spectrum}
The finite-length energy spectrum of the non-Hermitian SSH model exhibits a
$\mathcal{PT}$-symmetry-breaking phase transition in the non-Hermitian SSH model due to
the nonreciprocal hopping, manifested through the emergence of complex eigenenergy spectra. The other investigation we carry out is therefore what is the fate of the $\mathcal{PT}$ symmetry with a finite coupling to the cavity mode serving as the primary focus of our inquiry. Meanwhile, there exists a distinctive signature of its non-trivial topology, the presence of exponentially localized zero modes near the boundaries. Hence, we also investigate the behavior of the zero modes in the presence of cavity. To this extent we study the model within a mean field ansatz (see appendix A) assuming that there is no entanglement between the cavity ground modes $\left\vert \phi
\right\rangle $ and electronic ground state $\left\vert \varphi
\right\rangle $. The renormalised non-Hermitian SSH Hamiltonian dressed by
cavity obtained within this ansatz reads
\begin{equation}
		\begin{split}
			&H_{non-Hermitian-SSH}^{mf} =\left\langle \phi \right\vert H\left\vert \phi
			\right\rangle \\
			&=(v^{\prime }+\frac{\gamma ^{\prime }}{2}%
			)\sum_{j=1}^{L}c_{j,A}^{\dagger }c_{j,B}+(v^{\prime }-\frac{\gamma ^{\prime }%
			}{2})\sum_{j=1}^{L}c_{j,B}^{\dagger }c_{j,A}  \\
			&-w^{\prime }\sum_{j=1}^{L-1}c_{j+1,A}^{\dagger }c_{j,B}-w^{\prime
			}\sum_{j=1}^{L-1}c_{j,B}^{\dagger }c_{j+1,A},\\  \label{6}
		\end{split}
\end{equation}%
where $v^{\prime }=v\xi (g_{coupling},b_{0})$, $\gamma ^{\prime }=\gamma \xi (g_{coupling},b_{0})$
and $w^{\prime }=w\xi (g_{coupling},1-b_{0})$ are the dressed hoppings strength by
cavity with the factor

\begin{equation}
	\xi (g_{coupling},l)=\left\langle \phi \right\vert e^{-i\frac{g_{coupling}}{\sqrt{L}}%
		l(a+a^{\dagger })}\left\vert \phi \right\rangle .  \label{7}
\end{equation}%
Meanwhile, a renormalised photon Hamiltonian $H_{phot}$ is obtained within
the mean field ansatz (see Methods for the specific expression). By
self-consistently solving the corresponding mean-field Hamiltonian, we
obtain the photonic ground state $\left\vert \phi \right\rangle $ and the
electronic ground state $\left\vert \varphi \right\rangle $, and find that
the renormalization factor $\xi (g_{coupling},l)$ is only real due to the lack of a
finite current in the ground state of electron-photon coupling Hamiltonian.
It's worth noting that the numerical methods employed in obtaining\ the
ground state achieve a high level of accuracy, limited only by the cutoff of
the maximum photon number in the Fock space $N_{\max }^{phot}$, and we
choose $N_{\max }^{phot}=60$ for accuracy and simplicity.
\begin{figure*}
	\label{Fig.8} \centering\includegraphics[width=0.625%
	\columnwidth]{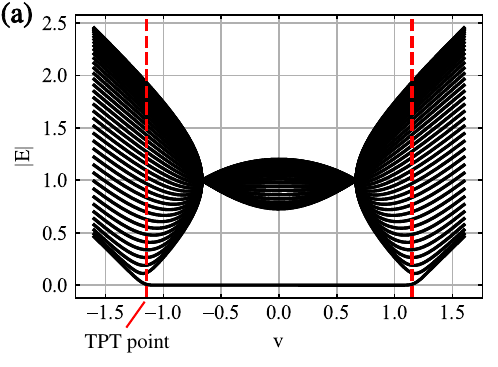}\includegraphics[width=0.65%
	\columnwidth]{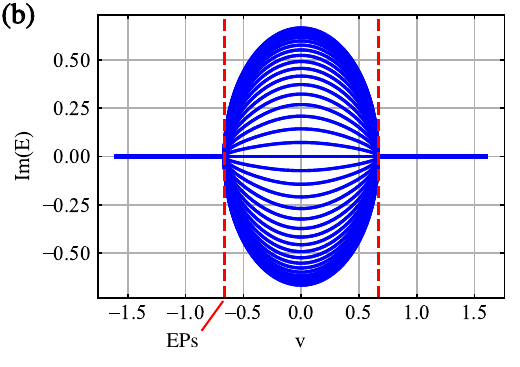}\includegraphics[width=0.62%
	\columnwidth]{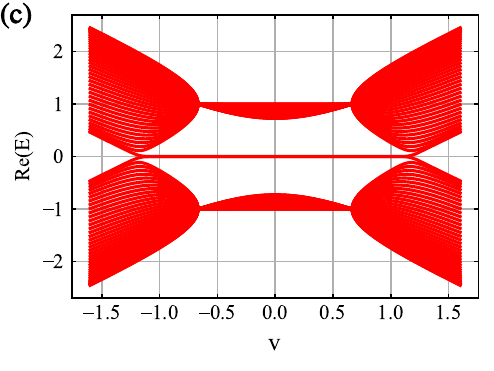}\newline
	\includegraphics[width=0.625\columnwidth]{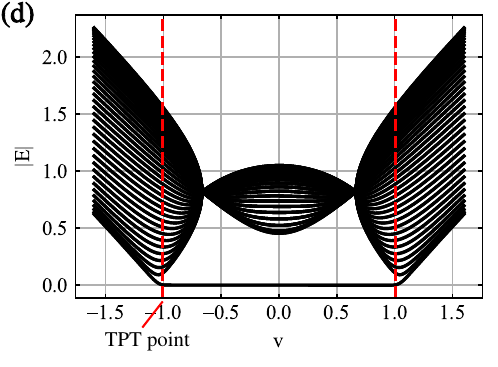}%
	\includegraphics[width=0.65\columnwidth]{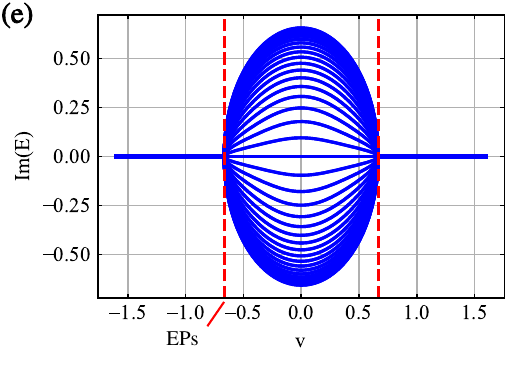}%
	\includegraphics[width=0.62\columnwidth]{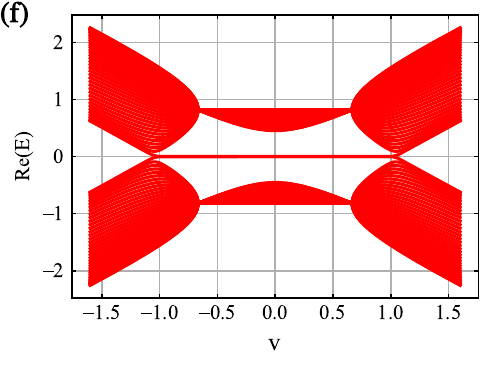}\newline
	\includegraphics[width=0.625\columnwidth]{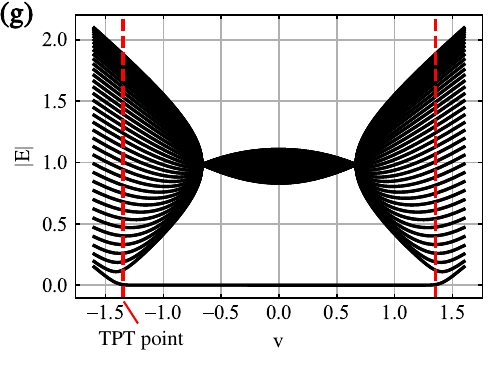}%
	\includegraphics[width=0.64\columnwidth]{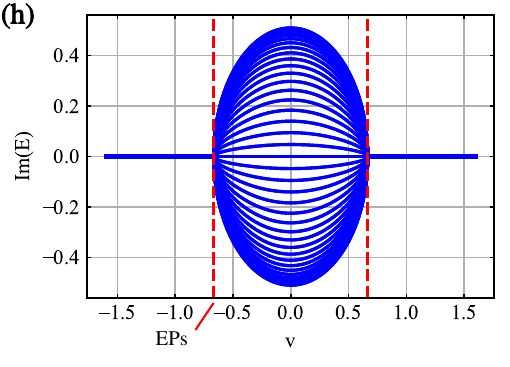}%
	\includegraphics[width=0.62\columnwidth]{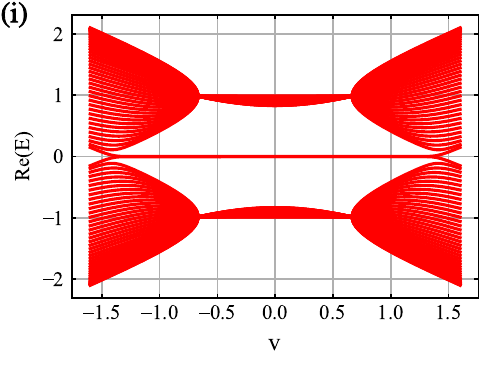}
	\caption{Energy spectrum of the non-Hermitian SSH model
		coupled to a single mode cavity as a function of lattice intracell hopping
		strength $v$. (a) the amplitude $\left\vert E\right\vert $, (b) imaginary Im$%
		(E)$ and (c) real Re$(E)$ part of energy spectrum in the absence of coupling
		to cavity $(g_{coupling}=0)$. (d) the amplitude $\left\vert E\right\vert $, (e)
		imaginary Im$(E)$ and (f) real Re$(E)$ part of energy spectrum in the
		presence of coupling to cavity $(g_{coupling}=10)$ with the intracell distance $%
		b_{0}=0.2$. (g) the amplitude $\left\vert E\right\vert $, (h) imaginary Im$%
		(E)$ and (i) real Re$(E)$ part of energy spectrum in the presence of
		coupling to cavity $(g_{coupling}=10)$ with the intracell distance $b_{0}=0.8$. Other
		parameters are chosen as intercell hopping stength $w=1$, nonreciprocal
		intercell hopping stength$\ \protect\gamma =4/3$, number of unit cells $L=40$%
		, lattice constant $a_{0}=1$, cavity frequency $\protect\omega _{c}=0.15$,
		and maximum photon number in the Fock space $N_{\max }^{phot}=60$.}
\end{figure*}
The finite-length energy spectrum of the non-Hermitian SSH model can be
obtainde by orthogonalizing $H_{non-Hermitian-SSH}$ and is plotted in Fig.
2(a)-(i) as functions of lattice intracell hopping amplitude $v$ for
different values of the light-matter coupling strength $g_{coupling}$ and intracell
distance $b_{0}$. As shown in Fig. 2(a), consistent with expectations for $%
g_{coupling}=0$, we observe a TPT when $v\approx \pm 1.2$ (red dashed line). During
this transition, a pair of exponentially small almost zero modes emerge
within the central region of the gap in the bulk in Fig. 2(c). Meanwhile,
there exists $\mathcal{PT}$-symmetry phase transition at EPs $(v\approx
\pm 0.6)$ which leads to emergence of imaginary energy spectrum as depicted
in Fig. 2(b). At $\mathcal{PT}$ symmetric region, the real part of energy spectrum in
Fig. 2(c)\ can captures the properties of non-Hermitian SSH model. Upon
considering light-matter coupling, the energy spectrum changes for different
intracell distance $b_{0}$. Especially, the TPT point(red dashed line) is
shifted to smaller values of $v$ for $b_{0}=0.2$ as shown in Fig. 2(d),
revealing that the TPT of non-Hermitian SSH model can be modified by the
quantum fluctuation of the cavity mode. In additional, In Fig. 2(g), it is
evident that the TPT point(red dashed line) is pushed to larger values of $v$
when $b_{0}=0.8$, and zero modes edge states can appear over a wider value
range of $v$ compared to the situation when $b_{0}=0.2$. This indicates that coupling to the quantum fluctuation of the light field can affect the TPT in the open boundary condition and this is consistent with the situation in the periodical boundary condition (see appendix B). In addition, the photonic spectral function's behavior which encoding how the light emitted from the cavity also show the TPT in the presence of cavity (see appendix C). However, unlike TPT point, EPs remain stable for different value of $b_{0}$ in Fig. 2(e) and Fig. 2(h), indicating the robustness of $\mathcal{PT}$-symmetry phase transition to
quantum fluctuation of cavity mode. However, the ground state of cavity will experience a sharp change when the $\mathcal{PT}$ symmetry of the non-Hermitian SSH model is broken, as we are going to discuss below.\

\section{The generation of\ squeezed displaced\ Schr\"{o}dinger-cat states}
\label{generation}
As is known to all, the photon groundstate of a pure cavity is vaccum state
where there only exists vacuum fluctuations without any photon. However,
different photon groundstate will emerge by considering the interaction
between an one-dimension electronic chain and cavity mode. Hence, we
investigate the groundstate of the electron-dressed cavity and show that the
 SDSc states can be obtained with
high fidelity when there exists a $\mathcal{PT}$-symmetry phase transition in
non-Hermitian SSH chain. To this extent we study the full Hamiltonian within a mean
field ansatz assuming that there is no correlation between the photon
ground state $\left\vert \phi \right\rangle $ and electronic ground state $%
\left\vert \varphi \right\rangle $. Thus, the photonic mean Hamiltonian
reads
\begin{widetext}
\begin{equation}
		\begin{split}
			H_{phot} =\left\langle \varphi \right\vert H\left\vert \varphi
			\right\rangle 
			=(v+\frac{\gamma }{2})e^{i\frac{g_{coupling}}{\sqrt{L}}b_{0}(a+a^{\dagger
				})}D_{AB}^{jj}+(v-\frac{\gamma }{2})e^{-i\frac{g}{\sqrt{L}}%
				b_{0}(a+a^{\dagger })}(D_{AB}^{jj})^{\dagger }\\
			-we^{i\frac{g_{coupling}}{\sqrt{L}}(1-b_{0})(a+a^{\dagger })}D_{AB}^{j+1,j}-we^{-i%
				\frac{g_{coupling}}{\sqrt{L}}(1-b_{0})(a+a^{\dagger })}(D_{AB}^{j+1,j})^{\dagger }
			+\omega _{c}(a^{\dagger }a+\frac{1}{2}), \label{8}
		\end{split}
\end{equation}%
\end{widetext}
where $D_{AB}^{jj}=$ $\sum_{j=1}^{L}\left\langle \varphi \right\vert
c_{j,A}^{\dagger }c_{j,B}\left\vert \varphi \right\rangle $ and $%
D_{AB}^{j+1,j}=$ $\sum_{j=1}^{L-1}\left\langle \varphi \right\vert
c_{j+1,A}^{\dagger }c_{j,B}\left\vert \varphi \right\rangle $ are the
hopping matrix element in the electronic ground state $\left\vert \varphi
\right\rangle $. By calculating the eigenvectors of the photonic Hamiltonian
$H_{phot}$, we find the photon ground state $\left\vert \phi \right\rangle $%
. To see the quantum features of the photon ground states, we now calculate
the Wigner function. The Wigner function, as a phase-space quasiprobability
distribution [3, 55], is defined in the $(q,p)$ position and momentum space
as
\begin{equation}
	W(q,p)=\frac{1}{2\pi }\int_{-\infty }^{\infty }dx\left\langle q-\frac{x}{2}%
	\right\vert \rho \left\vert q+\frac{x}{2}\right\rangle e^{ipx},  \label{9}
\end{equation}%
where $\rho $ is the density matrix of the photon state $\left\vert \phi
\right\rangle $.

As shown in Fig. 3(a) and Fig. 3(b), we plot the Wigner function of the
photon ground state $\left\vert \phi \right\rangle $ and the exact SDSc
state $\left\vert \phi ^{SDSc}\right\rangle =S(r)D(\lambda )(\frac{%
	\left\vert \alpha \right\rangle +\left\vert -\alpha \right\rangle }{\sqrt{2}}%
)$ in the $\mathcal{PT}$-symmetry phase of non-Hermitian SSH model $(v<\frac{%
	\gamma }{2})$. $S(r)=\exp [(ra^{\dagger 2}-r^{\ast }a^{2})/2]$ is the
squeeze operator, $D(\lambda )=\exp (\lambda a^{\dagger }-\lambda ^{\ast }a)$
is the displacement operator and $\frac{\left\vert \alpha \right\rangle
	+\left\vert -\alpha \right\rangle }{\sqrt{2}}$ is the Schr\"{o}dinger-cat
states. By comparing fig. 3(a) with fig. 3(b), we find that the phase-space
quasiprobability distribution of $\left\vert \phi \right\rangle $ and $%
\left\vert \phi ^{SDSc}\right\rangle $ are almost the same. Similar to Schr%
\"{o}inger's cat paradox, the presence of two peaks in the Wigner function
in fig. 3(a) and fig. 3(b) corresponds to both the \textquotedblleft dead
cat\textquotedblright\ and the \textquotedblleft alive
cat\textquotedblright\ states, and the observation of negative values in the
Wigner function signifies the nonclassical feature in SDSc state. We define
the fidelity $F$ between the photon ground (GS) states $\left\vert \phi
\right\rangle $\ in our system and the exact Squeezed Displaced\ Schr\"{o}%
dinger-cat states\ state $\left\vert \phi ^{SDSc}\right\rangle $ as
\begin{equation}
	F=\text{Tr}[\sqrt{\sqrt{\rho _{_{SDSc}}}\rho _{GS}\sqrt{\rho _{_{SDSc}}}}],
	\label{10}
\end{equation}%
where $\rho _{_{SDSc}}(\rho _{_{GS}})$ is the density matrix of the photon
state $\left\vert \phi \right\rangle (\left\vert \phi ^{SDSc}\right\rangle )$%
. As the fidelity $F$ approaches $1$, it indicates that the state generated
by our system closely resembles the ideal state, show a high degree of
similarity. After the numerical calculation, we find that the fidelity $F$
between the photon ground state $\left\vert \phi ^{GS}\right\rangle $ and
the exact SDSc state $\left\vert \phi ^{SDSc}\right\rangle $ is $0.994$
revealing that the cavity ground state is steered into a SDSc state with
high similarity.

\begin{figure}[ht!]
	\label{Fig.2} \centering\includegraphics[width=0.52%
	\columnwidth]{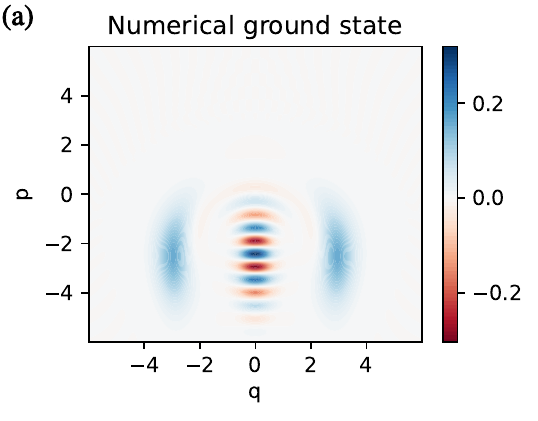}\includegraphics[width=0.52%
	\columnwidth]{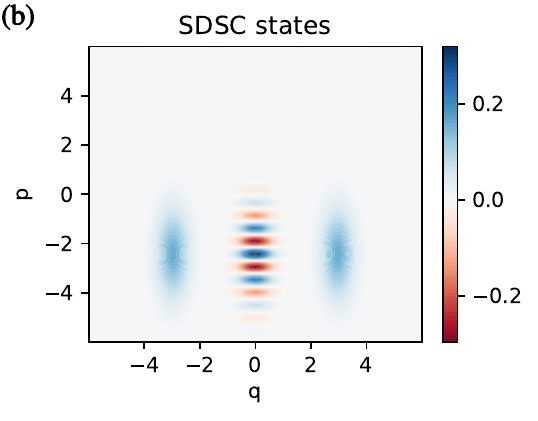}
	\caption{Wigner function for (a) the photon ground state $\left\vert \protect%
		\phi \right\rangle $ and (b) the exact SDSc state $\left\vert \protect\phi %
		^{SDSc}\right\rangle =S(r)D(\protect\lambda )(\frac{\left\vert \protect%
			\alpha \right\rangle +\left\vert -\protect\alpha \right\rangle }{\protect%
			\sqrt{2}})$ where $r=0.54$, $\protect\lambda =-i$ and $\protect\alpha =3.6$.
		Other parameters are chosen as intracell hopping stength $v=1.3$, intercell
		hopping stength $w=1$, nonreciprocal intercell hopping stength$\ \protect%
		\gamma =8/3$, number of unit cells $L=185$, intracell distance $b_{0}=0.9$,
		lattice constant $a_{0}=1$, cavity frequency $\protect\omega _{c}=0.15$,
		coupling strength $g_{coupling}=11.1$, and maximum photon number in the Fock space
		$N_{\max }^{phot}=60$.}
\end{figure}

However, the fidelity $F$ approaches $0$ when the $\mathcal{PT}$ symmetry of
non-Hermitian SSH model recovers $(v\geq \frac{\gamma }{2})$. We
showing this by plotting the fidelity $F$ as functions of coupling strength $%
g_{coupling}$ and the intracell distance $b_{0}$ for different values of
intracell hopping strength $v$ in Fig. 4(a)-(c). Fig. 4(a) shows a high
fidelity in the area of large value $b_{0}$, with a region of low fidelity
in the remaining area of examined parameter when the $\mathcal{PT}$ symmetry in non-Hermitian SSH chain is broken. Futhermore, from Fig. 4(a) we find that
the SDSc state $\left\vert \phi ^{SDSc}\right\rangle $ with high fidelity
can be obtained with weaker coupling strength $g_{coupling}$ as the value of $%
b_{0}$ increased. When the $\mathcal{PT}$ symmetry recovers $(v\geq \frac{\gamma }{2%
})$ the fidelity $F$ approaches $0$ although there exists maximum value of
the fidelity $F$ in Fig. 4(a) and 4(b). This implies a dramatic change in
the ground state of the cavity $\left\vert \phi ^{GS}\right\rangle $ in the
presence of $\mathcal{PT}$ symmetry of non-Hermitian SSH chain. We further show this by
plotting the fidelity $F$ in Fig. 4(d) as a function of intracell hopping
strength $v$ at fixed values of coupling strength $g_{coupling}$ and the
intracell distance $b_{0}$ depicted. It can be vividly seen from Fig. 4(d)
that the fidelity $F$ approaches $1$ as the $\mathcal{PT}$ symmetry is broken $(v<¦Ã
/2)$, undergoes an abrupt change at the EPs $(v=¦Ã /2)$, and
ultimately decreases to $0$ with the recover of the $\mathcal{PT}$ symmetry. This implies that the emergency of the SDSc state can be regarded as a signature of the $\mathcal{PT}$-symmetry phase transition in the non-Hermitian SSH model. It is worth to note that the ground state of the cavity is no longer the SDSc state but a cavity state with the vanishing of two peaks and negative values in Wigner function with the recover of the $\mathcal{PT}$ symmetry as depicted in the insets and it is an exact displaced squeezed vacuum state in the thermodynamic limit (see appendix D).

\begin{figure*}
	\label{Fig.3} \centering\includegraphics[width=0.63%
	\columnwidth]{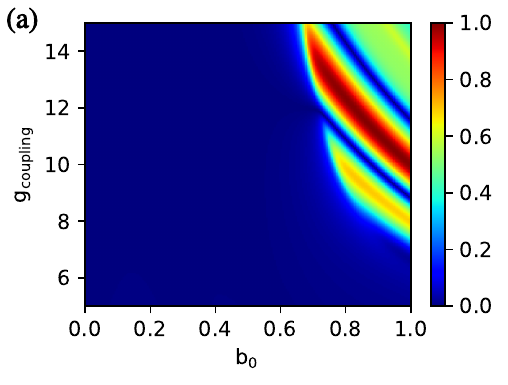}\includegraphics[width=0.68%
	\columnwidth]{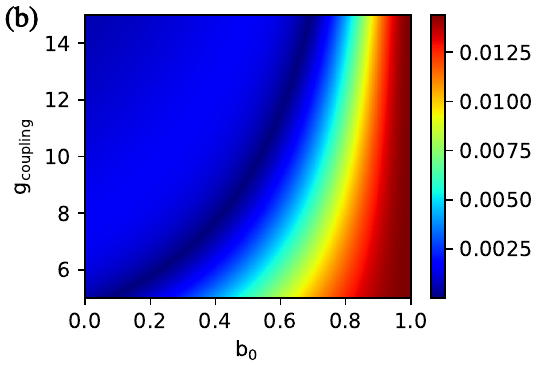}\includegraphics[width=0.66%
	\columnwidth]{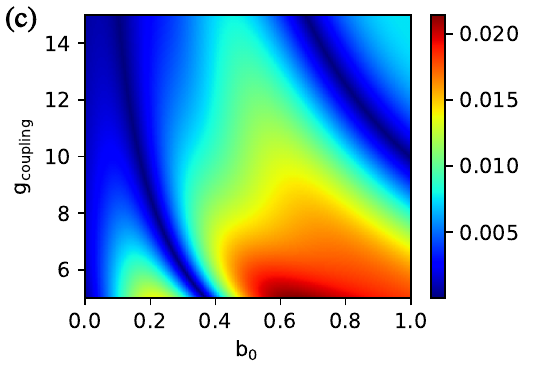}
	\centering\includegraphics[width=1.4%
	\columnwidth]{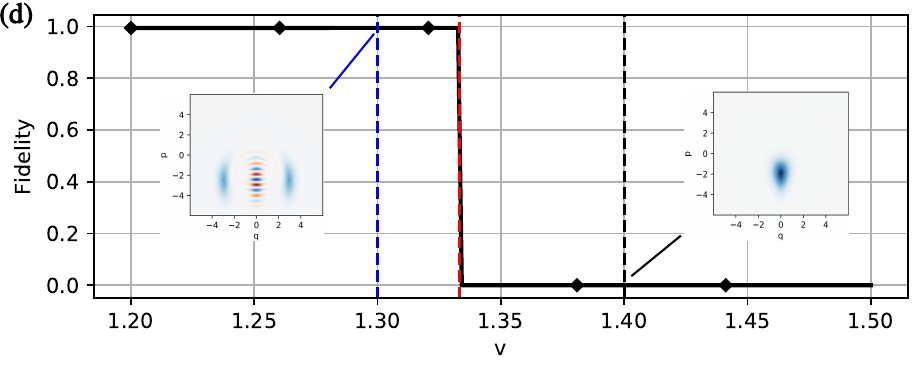}
	\caption{Fidelity $F$ between the photon ground (GS) states $\left\vert
		\protect\phi ^{GS}\right\rangle $ and the exact squeezed displaced\ Schr\"{o}%
		dinger-cat states\ state $\left\vert \protect\phi ^{SDSc}\right\rangle $ as
		functions of coupling strength $g_{coupling}$ and intracell distance $b_{0}$
		for (a) $v=1.3$, (b) $v=\protect\gamma /2$ and (c) $v=1.4$. (d) Fidelity $F$
		as functions of $v$ when $g_{coupling}=11.1$ and $b_{0}=0.9$, while the inset
		shows the Wigner function of cavity ground state at $v=1.3$ (blue dashed
		line) and $v=1.4$ (black dashed line). The red dashed line in (d) represents
		EPs $(v=\protect\gamma /2)$. Other parameters are chosen as intercell
		hopping stength $w=1$, nonreciprocal intercell hopping stength$\ \protect%
		\gamma =8/3$, number of unit cells $L=185$, lattice constant $a_{0}=1$,
		cavity frequency $\protect\omega _{c}=0.15$ and maximum photon number in the
		Fock space is $N_{\max }^{phot}=60$.}
\end{figure*}

The cavity ground state corresponds to the state with the lowest eigenvalue
of the electron-dressed cavity Hamiltonian. By examining the distribution of
the Hamiltonian in phase space, we can gain insight into the physical
mechanism underlying the alteration of the cavity ground state's properties.
The cavity Hamiltonian can be conveniently described as a harmonic
oscillator, allowing us to represent it using dimensionless position and
momentum operators in the semiclassical limit [] $x\equiv (a^{\dagger }+a)/2$
and $p\equiv (a^{\dagger }-a)/2i$. Thus, the semiclassical limit of Eq. (\ref%
{6}) can be rewritten as
\begin{equation}
		\begin{split}
			&H_{\mathrm{eff}}(x,p) =\left\langle \varphi \right\vert H\left\vert \varphi
			\right\rangle \\
			&=(v+\frac{\gamma }{2})e^{i\frac{g_{coupling}}{\sqrt{L}}%
				b_{0}2x}D_{AB}^{jj}+(v-\frac{\gamma }{2})e^{-i\frac{g_{coupling}}{\sqrt{L}}%
				b_{0}2x}(D_{AB}^{jj})^{\dagger } \\
			&-we^{i\frac{g_{coupling}}{\sqrt{L}}(1-b_{0})2x}D_{AB}^{j+1,j}-we^{-i\frac{g_{coupling}}{\sqrt{L}}%
				(1-b_{0})2x}(D_{AB}^{j+1,j})^{\dagger } \\
			&+\omega _{c}(x^{2}+p^{2}+\frac{1}{2}),\\ \label{11}
		\end{split}
\end{equation}
where $(x,p)\in R^{2}$ are now continuous (nonquantized) classical
variables. Therefore, the model described by a 1D tight-binding
non-Hermitian SSH model coupled to a single mode cavity exhibits a
semiclassical analogy with a single effective degree of freedom, and its
phase space is denoted as $M=R^{2}$. 
\begin{figure*}
	\label{Fig.4} \centering\includegraphics[width=0.62%
	\columnwidth]{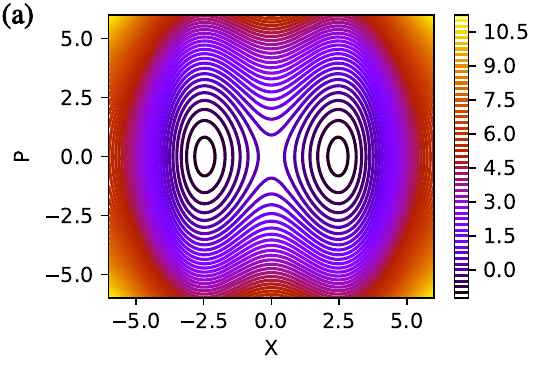}\includegraphics[width=0.6%
	\columnwidth]{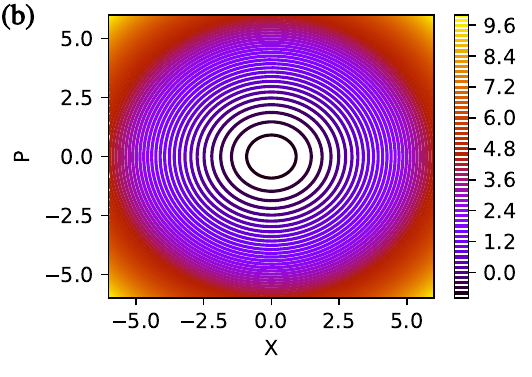}\includegraphics[width=0.62%
	\columnwidth]{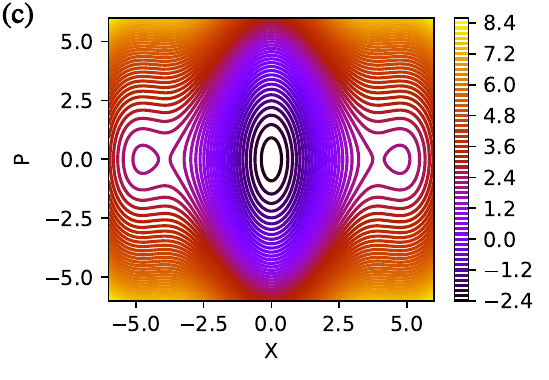}
	\centering\includegraphics[width=0.58%
	\columnwidth]{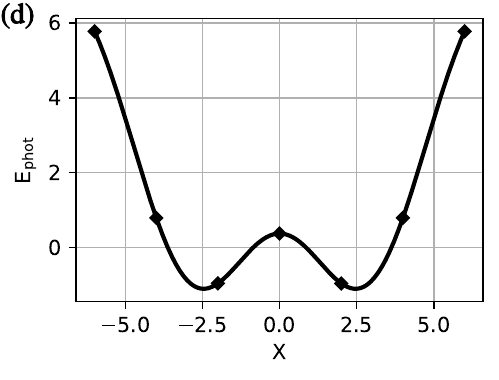}\includegraphics[width=0.6%
	\columnwidth]{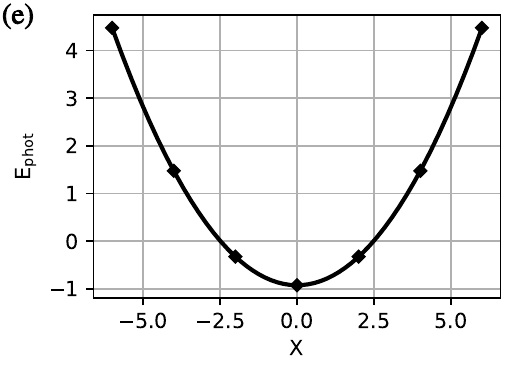}\includegraphics[width=0.6%
	\columnwidth]{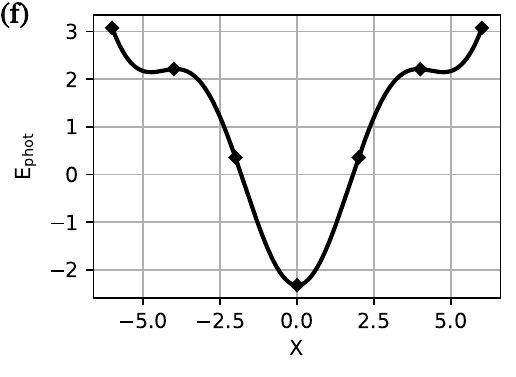}
	\caption{Constant energy contours of the full classical Hamiltonian as
		functions of $x$ and $p$ for (a) $v=1.3$, (b) $v=\protect\gamma /2$ and (c) $%
		v=1.4$. Constant energy contours of the full classical Hamiltonian as
		functions of $x$ when $p=0$ for (d) $v=1.3$, (e) $v=\protect\gamma /2$ and
		(f) $v=1.4$. Other parameters are chosen as intercell hopping stength $w=1$,
		nonreciprocal intercell hopping stength$\ \protect\gamma =8/3$, number of
		unit cells $L=185$, lattice constant $a_{0}=1$, cavity frequency $\protect%
		\omega _{c}=0.15$, coupling strength $g_{coupling}=11.1$ and maximum photon number in the Fock space is $N_{\max
		}^{phot}=60$.}
\end{figure*}
Mean-field properties of the quantum
model Eq. (\ref{9}) such as, e.g., the ground-state energy, the photon
population can be captured by studying the distribution of energy of
semiclassical Hamitonian. Especially, the emergence of SDSC state $%
\left\vert \phi ^{SDSc}\right\rangle $ can be understood through the
structure of the classical phase space. In Fig. 5(a), 5(b) and 5(c), we show
the energy contours of Hamitonian Eq. (\ref{9}), while the slice figure $%
(p=0)$ of energy are shown in Fig. 5(d), 5(e) and 5(f), accordingly. In Fig.
5(a), we observe that $H_{\mathrm{eff}}(x,p)$ has global extrema at nonzero values of $x$ and $p$. The classical phase space structure undergoes a complete
transformation when $\mathcal{PT}$ symmetry is broken $(v<\gamma /2)$, revealing the emergence of two minima corresponding to the ground-state energy. These minima are positioned at symmetric values relative to $x=0$ can be specifically shown in Fig. 5(d), and it makes the ground state of the cavity
exhibits a distribution in phase space similar to that of a SDSc state\
which is actually the two peaks in the Wigner function in Fig. 3(b).
However, as shown in Fig. 5(b) and 5(e) when $v=\gamma /2$, $H_{\mathrm{eff}}(x,p)$
allows for a single, global minimum at $x=p=0$, corresponding to the
ground-state energy. Although the energy of $H_{\mathrm{eff}}(x,p)$ in Fig. 5(c) and
5(f) has local extrema at nonzero values of $x$ and $p$, it has no
connection with the ground-state energy and has no effect on cavity ground
state. Therefore, one can see that phase-space quasiprobability distribution
of cavity ground state in the presence of $\mathcal{PT}$ symmetry has no two-peaks
compared to that of SDSc state due to the single, global minimum of energy
at $x=p=0$ when $v\geq \gamma /2$.

\section{Phase estimation with cavity ground state}
\label{Phase}
The ground state, as the lowest energy eigenstate of the system, can be
prepared as stable a quantum resource for quantum metrology without
excitating a system to high energy level. In non-Hermitian systems, the $\mathcal{PT}$ symmetry of the system may influence the ground state, as previously
discussed, thereby impacting the quality of the ground state as a quantum
resource. To this extend, we utilize the generated cavity ground states to
estimate the phase in the optical interferometer and investigate the Quantum
Fisher Information (QFI). Furthermore, the nonclassicality measured using an
operational resource theory (ORT)[] will be discussed.

As depicted in Fig. 6, the optical interferometer comprises two linear phase
shifters, incorporating an unknown relative phase $\Phi _{-}=(\Phi _{1}-\Phi
_{2})$ between the arms. The phase information will be encoded within the
initial quantum state (cavity ground state $\left\vert \phi
^{GS}\right\rangle $) at the input of the interferometer through unitary
evolution, i.e.,
\begin{equation}
	\left\vert \Psi \right\rangle =U_{phase-shift}\left\vert \Psi
	_{0}\right\rangle ,  \label{12}
\end{equation}

where $U_{phase-shift}=\exp [i(\Phi _{1}a_{1}^{\dagger }a_{1}+\Phi
_{2}a_{2}^{\dagger }a_{2})]$, and $a_{1,2}^{\dagger }(a_{1,2})$ represent
annihilation (creation) operators in the arms 1 and 2. $\left\vert \Psi
_{0}\right\rangle =\left\vert \phi ^{GS}\right\rangle \otimes \left\vert
\phi ^{GS}\right\rangle $ is the input state of the interferometer. Here, we
concentrate on the relative phase between the two arms.

\begin{figure}[ht!]
	\label{Fig.5} \centering\includegraphics[width=0.9\columnwidth]{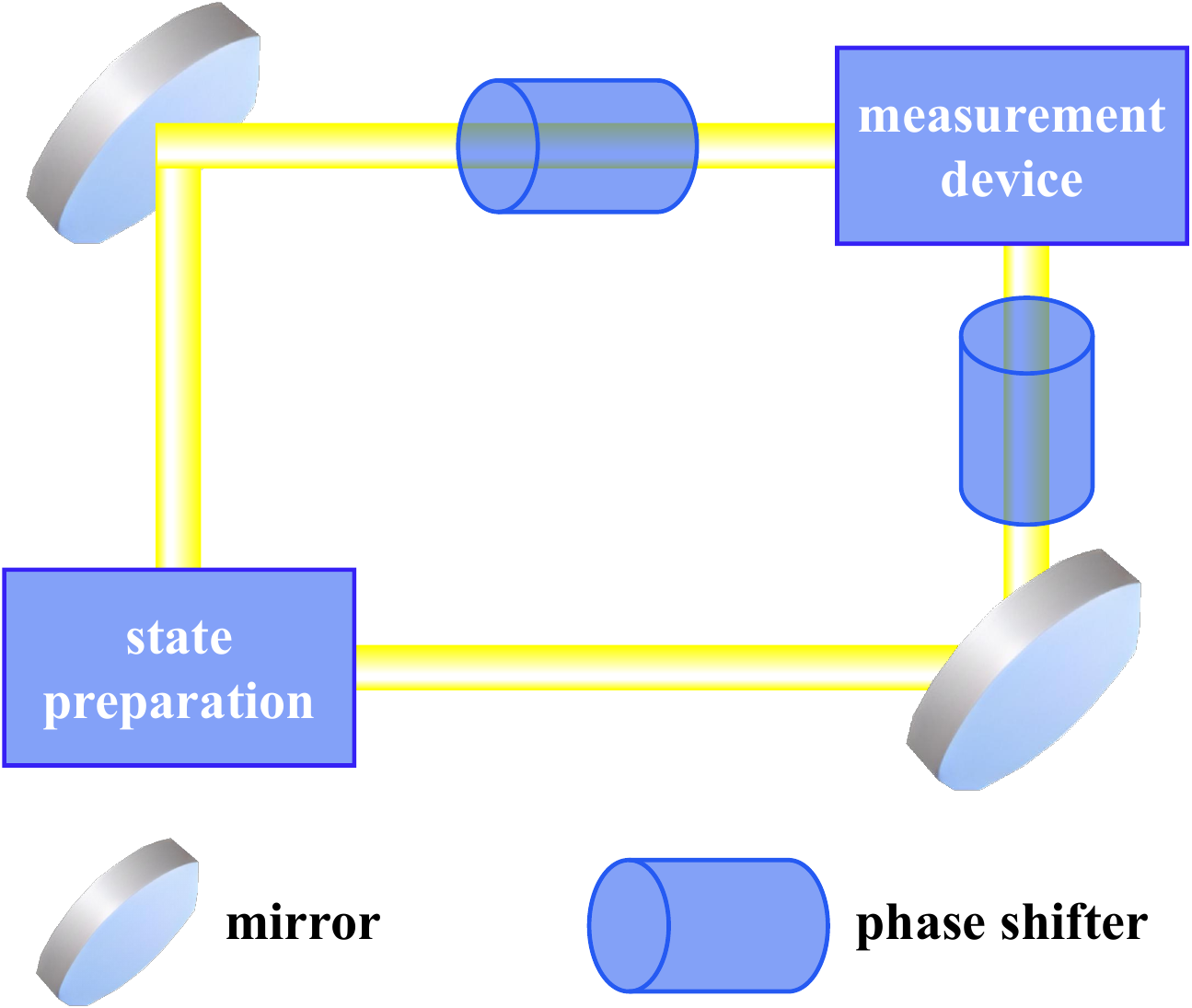}
	\caption{An optical interferometer model is utilized for phase estimation. A
		prepared quantum state is input into the interferometer, where an unknown
		relative phase $(\Phi _{1}-\Phi _{2})$ is introduced by two linear phase
		shifters between its arms. The phase information becomes encoded in the
		initial quantum state and is subsequently measured at the output ports.
		Other parameters are chosen as intercell hopping stength $w=1$, intracell
		hopping stength $v=1.3$, nonreciprocal intercell hopping stength$\ \protect%
		\gamma =8/3$, number of unit cells $L=185$, lattice constant $a_{0}=1$,
		cavity frequency $\protect\omega _{c}=0.15$, coupling strength $g_{coupling}=11.1$, and maximum photon number in the
		Fock space is $N_{\max }^{phot}=60$.}
\end{figure}
\begin{figure}[ht]
	\label{Fig.6} \centering\includegraphics[width=0.9%
	\columnwidth]{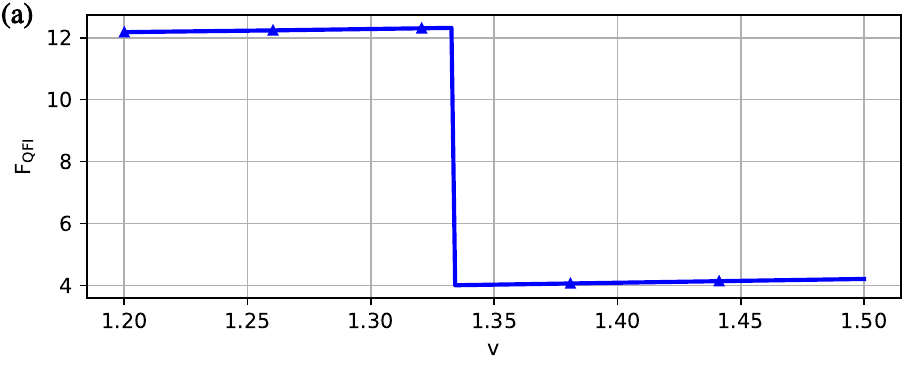}
	\centering
	\includegraphics[width=0.9%
	\columnwidth]{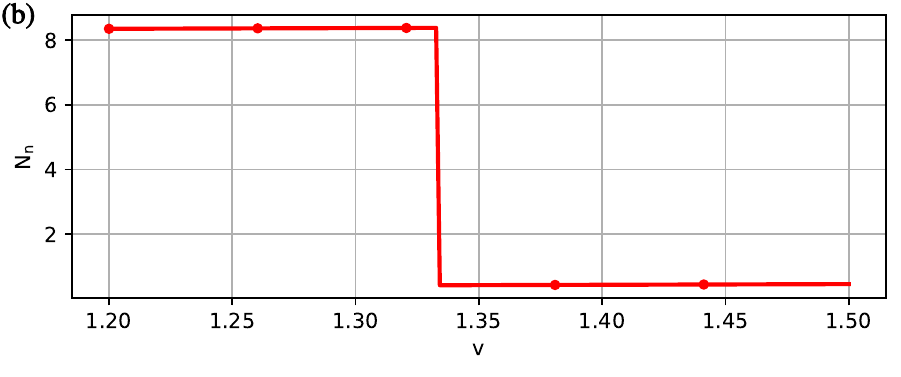}
	\caption{(a) QFI $F_{QFI}$ and (b) the nonclassicality $N_{n}$ as functions
		of intracell hopping stength $v$. Other parameters are chosen as intercell
		hopping stength $w=1$, nonreciprocal intercell hopping stength$\ \protect%
		\gamma =8/3$, $g_{coupl}=11.1$, $b_{0}=0.9$, number of unit cells $L=185$,
		lattice constant $a_{0}=1$, cavity frequency $\protect\omega _{c}=0.15$, coupling strength $g_{coupling}=11.1$, and
		maximum photon number in the Fock space is $N_{\max }^{phot}=60$.}
\end{figure}
Considering a pure input state with path-symmetric, the unitary evolution
can be simplified to $\exp [i\Phi _{-}(a_{1}^{\dagger }a_{1}-a_{2}^{\dagger
}a_{2})/2]$. Then, the QFI of the unitary evolution of the initial state $%
\left\vert \Psi _{0}\right\rangle $\ can be defined as[]

\begin{equation}
	F_{QFI}=4\left\langle \Psi _{0}\right\vert \Delta ^{2}H\left\vert \Psi
	_{0}\right\rangle ,  \label{13}
\end{equation}

where $H=i(\partial _{\Phi _{-}}U_{phase-shift}^{\dagger })U_{phase-shift}$
and $\Delta ^{2}H=(H-\left\langle H\right\rangle )^{2}$. With the given
operator $U_{phase-shift}$ expression, the QFI can be calculated as
\begin{equation}
	F_{QFI}=2[\text{Var}_{\Psi _{0}}(a_{i}^{\dagger }a_{i})-\text{Cov}_{\Psi
		_{0}}(a_{1}^{\dagger }a_{1},a_{2}^{\dagger }a_{2})],i=1\text{ or }2,
	\label{14}
\end{equation}
where Var$_{\Psi _{0}}(\cdot )$ and Cov$_{\Psi _{0}}(\cdot )$ are the
variance and the covariance in the input state $\left\vert \Psi
_{0}\right\rangle $, correspondingly. It can be clearly seen from Eq. (\ref%
{12}) that QFI only depends on the properties of the input state $%
(\left\vert \phi ^{GS}\right\rangle \otimes \left\vert \phi
^{GS}\right\rangle )$. Considering the initial state here is separable Cov$%
_{\Psi _{0}}(a_{1}^{\dagger }a_{1},a_{2}^{\dagger }a_{2})=0$, thus the QFI
can be reduced to $2$Var$_{\Psi _{0}}(a_{i}^{\dagger }a_{i})$. In addition,
a nonclassicality measure for quantum state rooted in the framework of ORT
has been introduced, demonstrating various essential properties as a
valuable resource for phase estimation. For pure states, nonclassicality
measured based on ORT reads

\begin{equation}
	N_{n}=\left\langle a^{\dagger }a\right\rangle -\left\vert \left\langle
	a\right\rangle \right\vert ^{2}+\left\vert \left\langle a\right\rangle
	-\left\langle a\right\rangle ^{2}\right\vert .  \label{15}
\end{equation}

In order to study the impact of $\mathcal{PT}$ symmetry on phase estimation utilizing
the cavity ground state as a quantum resource, we plot the QFI $F_{QFI}$ and
the nonclassicality $N_{n}$ as functions of intracell hopping stength $v$ in
Fig. 7(a) and 7(b), accordingly. From Fig. 7(a), a clear observation can be
made that the value of QFI $F_{QFI}$ is large as the $\mathcal{PT}$ symmetry is broken $%
(v<¦Ã /2)$ which indicates that the cavity ground state is an
appropriate quantum resource for phase estimation. However, there exists a
sudden transition in value of $F_{QFI}$\ at the exceptional point $(v=¦Ã
/2)$, and it suddenly become small as the $\mathcal{PT}$ symmetry recovers. Thus, the cavity ground state with $\mathcal{PT}$ symmetry in non-Hermitian SSH model is not
suitable for estimatin phase. From Fig. 7(b), it is evident that the value
of the nonclassicality $N_{n}$ experiences a significant shift at the EPs. Simultaneously, the cavity ground state with broken $\mathcal{PT}$-symmetry
exhibits higher nonclassicality, whereas the cavity ground state will lose
some nonclassicality in the presence of $\mathcal{PT}$ symmetry. Therefore, the ground
state of the cavity can be viewed as a good quantum resource for phase
estimation when $\mathcal{PT}$ symmetry in non-Herimtian SSH model is broken. \
\section{Conclusion}
\label{Conclusion}
In this work, we have investigated a non-Hermitian SSH chain coupled to a single spatially constant cavity mode. The open-boundary energy spectrum shows that topological phase transition in the non-Hermitian SSH chain can be modified by the fluctuation of cavity mode and captured in the photonic spectral function's
behavior. However, $\mathcal{PT}$-symmetry phase transition can not be modified in the presence of cavity and is difficult to detect. Fortunately, the emergency of SDSc state with high fidelity can capture $\mathcal{PT}$-symmetry phase transition in non-Hermitian SSH model coupled to the quantum light field of a cavity. When $\mathcal{PT}$ symmetry is broken the fidelity between exact SDSc state and cavity ground state approaches to almost 1, and then experiences a sharp drop to 0 as $\mathcal{PT}$ symmetry recovers. The emergency of SDSc state can naturally be understood by inspecting the effective Hamiltonian $H_{\mathrm{eff}}(x,p)$ in semiclassical limit: $H_{\mathrm{eff}}(x,p)$ has local extrema at two sides of x=0 when $\mathcal{PT}$ symmetry is broken. Therefore, our work therefore provides a convenient way to capture the $\mathcal{PT}$-symmetry phase transition in an electronic system. 
Besides, we exploit the ground state of cavity to estimate the phase in an optical interferometer, and find that QFI $F_{QFI}$ and the nonclassicality $N_{n}$ of the ground state will experience a sharp drop at the $\mathcal{PT}$-symmetry phase transition point. This indicates it is beneficial for quantum metrology and quantum information processing when considering the $\mathcal{PT}$-symmetry is broken in electronic system coupled to a cavity. Therefore, except for the flexible control and convenient probe of topological properties of the electronic systems provided by cavity photon, the ground state of cavity photon can be used to capture the $\mathcal{PT}$-symmetry phase transition in electronic system.

\appendix
\section{ Mean-field method to decouple electrons and photons in finite-length non-Hermitian SSH chain in the presence of cavity}
We present a detailed analysis of the solution to the light-matter Hamiltonian $H $ Eq. (\ref%
{5}) within the mean field approximation, which neglects correlations between cavity modes $\left\vert \phi
\right\rangle $ and electrons $\left\vert \varphi
\right\rangle $.
\begin{equation}
|\Psi\rangle=|\psi\rangle|\phi\rangle. \label{16}
\end{equation}
Due to the mean field decoupling, we are tasked with solving an electronic mean field Hamiltonian as described by Eq. (\ref
{6}), alongside a photonic mean field Hamiltonian characterized expressed as Eq. (\ref{8}). We determine the electronic spectrum of the non-Hermitian SSH model in the presence of coupling to cavity photons through the self-consistent solution of Eq. (\ref{6}) and Eq. (\ref{8}). Specifically, we fix g = 0 in $H_{non-Hermitian-SSH}^{mf}$ and numerically compute $D_{AB}^{jj}=$ $\sum_{j=1}^{L}\left\langle \varphi \right\vert
c_{j,A}^{\dagger }c_{j,B}\left\vert \varphi \right\rangle $ and $%
D_{AB}^{j+1,j}=$ $\sum_{j=1}^{L-1}\left\langle \varphi \right\vert
c_{j+1,A}^{\dagger }c_{j,B}\left\vert \varphi \right\rangle $. Then we insert
$D_{AB}^{jj}$ and $D_{AB}^{j+1,j}$ into $H_{phot}$ in Eq. (\ref{8}) with  $g \neq 0$, and evaluate $v^{\prime }$, $\gamma ^{\prime }$ and $w^{\prime }$. Subsequently, we insert $v^{\prime }$, $\gamma ^{\prime }$ and $w^{\prime }$ (computed for ($g \neq 0$)) into $H_{non-Hermitian-SSH}^{mf}$, then insert them into $H_{non-Hermitian-SSH}^{mf}$ to calculate $D_{AB}^{jj}$ and $D_{AB}^{j+1,j}$ again, and repeat the procedure until convergence. We find that the energy spectrum of the non-Hermitian SSH chain is modified in presence of cavity as shown in Fig. 2.

\section{TPT in Non-Bloch Energy spectrum}

In the electron lattices, the topological phase transition in the periodic
boundary condition would be associated with a closing and reopening of the
gap in the energy spectrum. Thus, we proceed to explore the energy
spectrum under
the periodic boundary condition, focusing on the occurrence of topological
phase transitions. Before this, we note that the usual Bloch waves carry a
pure phase factors $e^{ik}$ where $k$ is Bloch wave vector in Bloch
Brillouin zone. However, the role of the phase factors $e^{ik}$\ is now
played by $\beta =\sqrt{(v-\gamma /2)/(v+\gamma /2)}e^{ik}$ which has a
modulus $\left\vert \beta \right\vert \neq 1$ in general. This allow us to
write the Hamitonian in non-Bloch Brillouin zone(see Methods). With the
non-Bloch ferminoic field $\psi _{\beta }^{\dagger }=(c_{\beta ,A}^{\dagger
},c_{\beta ,B}^{\dagger })$,\ the renormalized Hamitonian under the periodic
boundary condition can be written in compact form as

\begin{equation}
	H=\sum_{\beta }\psi _{\beta }^{\dagger }\frac{h(\beta )}{i\beta }\psi
	_{\beta }+\omega _{c}(a^{\dagger }a+\frac{1}{2}),  \label{17}
\end{equation}%
where the non-Bloch Hamitonian $h(\beta )$ has this form
\begin{widetext}
	\begin{equation}
		\begin{split}
			h(\beta )=\left[ e^{-i\frac{g_{coupling}}{\sqrt{L}}b_{0}(a+a^{\dagger })}(v-\gamma
			/2)+e^{i\frac{g_{coupling}}{\sqrt{L}}(1-b_{0})(a+a^{\dagger })}\beta w\right] \sigma
			^{-}\\+\left[ e^{i\frac{g_{coupling}}{\sqrt{L}}b_{0}(a+a^{\dagger })}(v+\gamma /2)+e^{-i%
				\frac{g_{coupling}}{\sqrt{L}}(1-b_{0})(a+a^{\dagger })}\beta ^{-1}w\right] \sigma ^{+},
			\label{18}
		\end{split}
	\end{equation}%
\end{widetext}
where $\sigma ^{\pm }=(\sigma _{x}\pm i\sigma _{y})/2$ are pseudo-spin
up(down) operator. In fact, the Hamiltonian of such a bipartite lattice can
be written in the form of pseudo-spin components $h(\beta
)=\sum\limits_{i=x,y,z}d_{\beta i}\sigma _{i}$ by considering periodic
boundary conditions. Here, at the first glance, it is apparent that the
non-Bloch Hamiltonian $h(\beta )$ in Eq. (\ref{21}) lacks the presence of an
electronic mass term, i.e. $d_{\beta z}=0$, indicating the preservation of
chiral symmetry in the presence of cavity, i.e. $g\neq 0$. This guarantees
the existence of eigenstate pairs with opposite energies.

By applying the mean-field method described in the previous section(see
Methods), we perform calculations of the bulk energy spectrum while
considering the interaction between electron and photons. Through the
self-consistent solution of the electronic and photonic mean-field
Hamiltonian, we observe that the influence of photons on the electronic
hoppings which is characterized by purely real renormalization values $\xi
(g,l)$, so the electronic bulk energy spectrum reads
\begin{widetext}
\begin{equation}
	\begin{split}
	\varepsilon =\pm \sqrt{R_{+}R_{-}}
	=\pm \sqrt{(v^{\prime })^{2}-(\gamma
		^{\prime })^{2}/4+(w^{\prime })^{2}-2w^{\prime }\text{cos}k\cdot \sqrt{%
			(v^{\prime }+\gamma ^{\prime }/2)(v^{\prime }-\gamma ^{\prime }/2)}},
	\label{19}
	\end{split}
\end{equation}%
\end{widetext}
where $R_{\pm }=(v^{\prime }\pm \gamma ^{\prime }/2)+\beta ^{\mp 1}w^{\prime
}$ are the pseudo-spin up(down) components of the non-Bloch Hamitonian $%
h(\beta )$ with $v^{\prime }$, $\gamma ^{\prime }$ and $w^{\prime }$ defined
before Eq. (\ref{19}). It is worth noting that $\varepsilon $\ is real at
outside the EP point $(\left\vert v^{\prime }\right\vert >\left\vert \gamma
^{\prime }/2\right\vert )$.

Solving the photonic and electronic Hamiltonian self consistently we find
numerically the Non-Bloch energy spectrum that we plot for three values of
coupling strength $g$\ in Fig. 8 (panel (a) $g=5$, panel (b) $g=14$ and
panel (c) $g=20$). 
\begin{figure*}
	\label{Fig.9} \centering\includegraphics[width=0.6%
	\columnwidth]{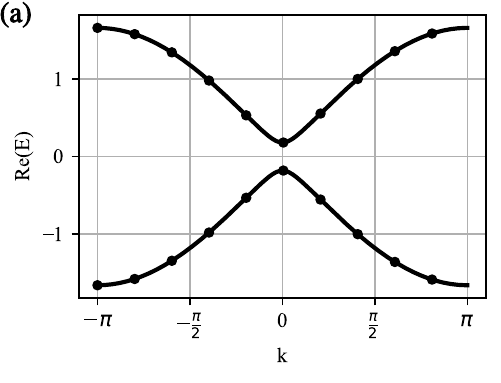}\includegraphics[width=0.6%
	\columnwidth]{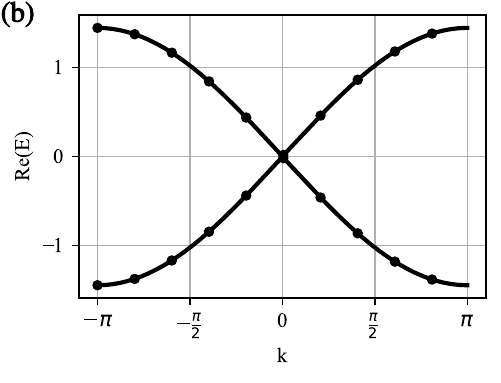}\includegraphics[width=0.6%
	\columnwidth]{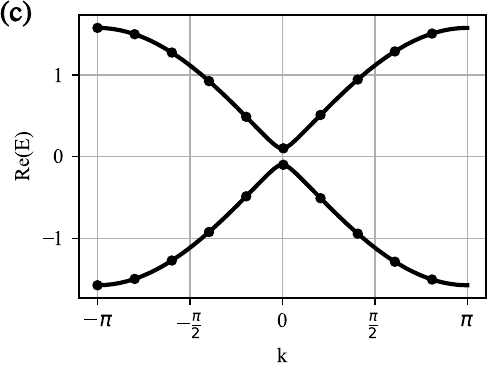}\newline
	\includegraphics[width=0.6\columnwidth]{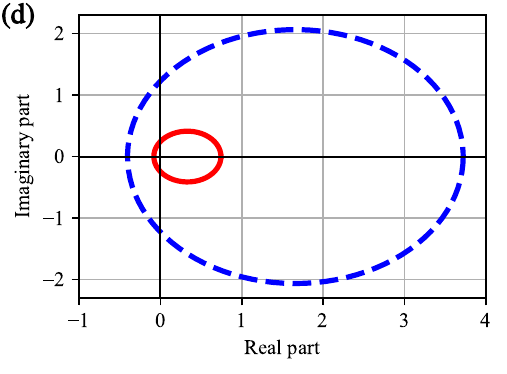}%
	\includegraphics[width=0.6\columnwidth]{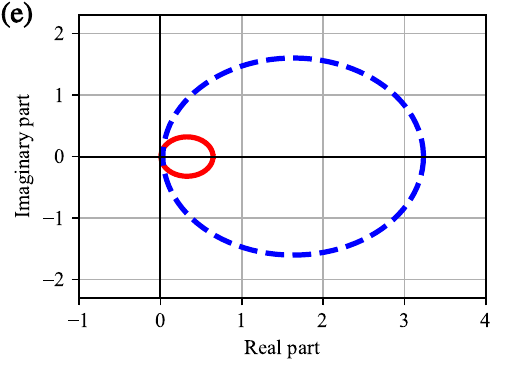}%
	\includegraphics[width=0.6\columnwidth]{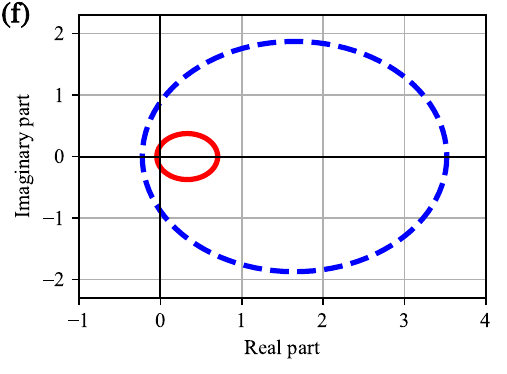}\newline
	\caption{Non-Bloch Energy spectrum $\protect\varepsilon $ in general
		brillouin zone with (a) $g=5,$(b) $g=14$ and (c) $g=20$ correspondingly.
		Trajectories of $R_{\pm }$ in $\protect\varepsilon $ with $k$ in $\protect%
		\beta $\ running from $0$ to $2\protect\pi $ with (a) $g=5,$(b) $g=14$ and
		(c) $g=20$ correspondingly. Other parameters are fixed as intercell hopping
		stength $w=1$, nonreciprocal intercell hopping stength$\ \protect\gamma =4/3$%
		, $b_{0}=0.2$, lattice constant $a_{0}=1$, cavity frequency $\protect\omega %
		_{c}=0.15$, and maximum photon number in the Fock space $N_{\max }^{phot}=60$%
		.}
\end{figure*}
It is clear that there exists eigenstate pairs with
opposite energies due to the preservation of chiral symmetry. Fig. 8(a) was
exhibited to demonstrate a frequency gap between two bands when $g=5$, but
Fig. 8(b) incidentally shows that the gap is closed at the center of the
Brillouin zone as $g=14$, where they touch at Dirac points. When $g$ is
increased to $20$, the gap is reopened with topological non-triviality which
can be seen in the trajectories of $R_{\pm }$ as shown in Fig. 8(d), (e) and
(f). Figures 8(d) show that $R_{\pm }$ have no wrapping to the origin and
thus the electron system is topological trivial as $g=5$. When $g$ is
increased to $14$, the trajectories of $R_{\pm }$ touch the origin and TPT
happens. Then an anticlockwise and a clockwise wrappings to the origin are
formed by $R_{+}$ and $R_{-}$, respectively, and thus the electron system is
topological non-trivial as $g$ increases across the value of $14$. This
gives a geometric picture to the TPT in Fig. 8(a), (b) and (c).

\section{Photonic spectrum with topological phase transition}

The excitaion of electron in the non-Hermitian SSH system coupled to a
cavity mode acts as hybrid light-matter polariton quasiparticles whose
energies are affected by the topological phase transition in the electronic
system. Meanwhile, the change in the properties of the polartion
quasiparticles can be read out from a photonic spectrum which also unveils
pivotal indications of the emission of the light from cavity. Thus, the
topological property of the electronic system can be captured from the
photonic spectrum. Particularly, we specifically investigate the photon
spectral function $A(\omega )=-\frac{1}{\pi }$Im$\int dte^{-i\omega
	t}(-i\theta (t))\left\langle [a(t),a^{\dagger }]\right\rangle $ influenced
by the electronic system, taking into account $1/L$ Gaussian fluctuations.
By using photon Green's function method[] $A(\omega )$ can be expressed as

\begin{equation}
	A(\omega )=-\frac{1}{\pi }\frac{\chi ^{\prime \prime }(\omega )(\omega
		+\omega _{c})^{2}}{(\omega ^{2}-\omega _{c}^{2}-2\omega _{c}\chi ^{\prime
		}(\omega ))^{2}+(2\omega _{c}\chi ^{\prime \prime }(\omega ))^{2}},
	\label{24}
\end{equation}%
where $\chi (\omega )=K(\omega )-\left\langle J_{d}\right\rangle $ is the
current-current correlation function comprises both paramagnetic and
diamagnetic contributions, and $\chi ^{\prime }(\omega )(\chi ^{^{\prime
		\prime }}(\omega ))$ are the real(imaginary) part of $\chi (\omega )$(see
Methods),

	\begin{equation}
		\begin{split}
			K(\omega ) =\int_{-\infty }^{\infty }\left\langle T_{c}J_{p}(\tau
			)J_{p}(0)\right\rangle e^{i\omega \tau }d\tau ,  \\
			J_{p} =\frac{g}{\sqrt{L}}\sum\limits_{\beta }\psi _{\beta }^{\dagger
			}(i\beta ^{-1}w\sigma ^{+}-i\beta w\sigma ^{-})\psi _{\beta }, \\
			J_{d} =-\frac{g^{2}}{\sqrt{L}}\sum\limits_{\beta }\psi _{\beta }^{\dagger
			}(\beta w\sigma ^{-}+\beta ^{-1}w\sigma ^{+})\psi _{\beta }, \label{25}
		\end{split}
	\end{equation}

where $J_{p}(\tau )$ and$\ J_{p}(0)$ are the current operators in the
Heisenberg picture.
\begin{figure*}
	\label{Fig.11} \centering\includegraphics[width=0.7%
	\columnwidth]{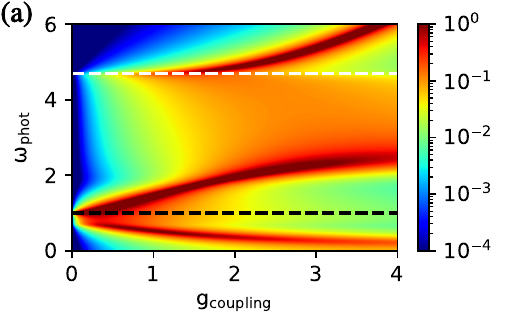}\includegraphics[width=0.7%
	\columnwidth]{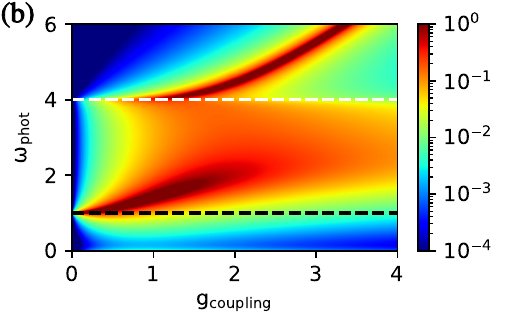}\includegraphics[width=0.7%
	\columnwidth]{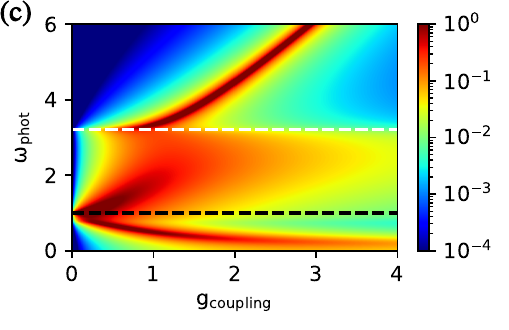}
	\caption{Photonic spectral function as a function of the light-matter
		coupling $g_{coupling}$ and photon frequency $\protect\omega _{phot}$. White
		dashed line corresponds to cavity frequecy $\protect\omega _{c}$ and black
		dashed line corresponds to $2E_{k=\protect\pi }$(the electronic bulk energy
		spectrum at $k=\protect\pi $ in Eq. (\protect\ref{10})). (a) intracell
		hopping $v=1.5$ and intercell hopping $w=1$ $(v>\protect\sqrt{(w)^{2}+(%
			\protect\gamma )^{2}/4})$ for topological trivial phase (b) $v=1.2$ and $w=1$
		$(v=\protect\sqrt{(w)^{2}+(\protect\gamma )^{2}/4})$ for TPT (c) $v=0.9$ and
		$w=1$ $(v<\protect\sqrt{(w)^{2}+(\protect\gamma )^{2}/4})$ for topological
		non-trivial phase. Other parameters are fixed as nonreciprocal intercell
		hopping stength$\ \protect\gamma =4/3$, lattice constant $a_{0}=1$, cavity
		frequency $\protect\omega _{c}=1$, and maximum photon number in the Fock
		space $N_{\max }^{phot}=60$.}
\end{figure*}
As depicted in Fig. 9, we illustrate the photonic spectral function's
behavior in the presence of cavity. As shown in Fig. 9(a), in the
topological trivial phase $(v>\sqrt{(w)^{2}+(\gamma )^{2}/4})$, there only
exists one peak in the photonic spectral function in the absence of cavity $%
(g_{coupling}=0)$, corresponding the cavity mode with frequency $\omega _{c}$
as depicted by the black dashed line. As the coupling strength $g_{coupling}$
increasing, the peak splits into two branches indicating the appearance of
the two hybrid light-matter polartion modes and the cavity mode with the
frequency higher than $2E_{k=\pi }$ (white dashed line) can hybrid with the
electron as shown in Fig. 9(b). As we fix the hopping parameters such that $v=\sqrt{%
	(w)^{2}+(\gamma )^{2}/4}$, the lowest branch of the photonic spectral
function disappear revealing the absence of the polartion modes with the
lowest frequency. This is directly linked to the closure of the bulk gap as
depicted in Fig. 8(b), serving as a clear indication of the topological
phase transition. When \ $v<\sqrt{(w)^{2}+(\gamma )^{2}/4}$, that is
topological non-trivial phase, the lowest branch of the photonic spectral
function reappear indicating the reemergence of the lowest frequency
polartion branch depicted in Fig. 9(c). Thus, the topological property of non-hermitian SSH chain can be probed by investigating the photonic spectral function's behavior in the presence of cavity. \

\section{Generation of displaced squeezed state }

In addition to being able to generate squeezed displaced\ Schr\"{o}%
dinger-cat states, other photonic\ states can emerge in the system where
non-Hermitian SSH chain coupled to a single mode cavity. Therefore, we focus on the emergency of the displaced squeezed vacuum states. We expand $H_{phot}$ in Eq. (\ref{8}) to second order in the field that gave the only non-vanishing contribution in the thermodynamic limit to the ground state energy. It reads
\begin{widetext}
\begin{equation}
	H_{phot}^{2nd}=\omega _{c}(a^{\dagger }a+\frac{1}{2})+\frac{g_{coupling}}{\sqrt{L}}%
	(a+a^{\dagger })\Pi -\frac{g_{coupling}^{2}}{2L}(a+a^{\dagger })^{2}\Lambda  \label{22}
\end{equation}%
\end{widetext}
where $\Lambda =(v+\frac{\gamma }{2})D_{AB}^{jj}b_{0}^{2}+(v-\frac{\gamma }{2%
})(D_{AB}^{jj})^{\dagger
}b_{0}^{2}-wD_{AB}^{j+1,j}(1-b_{0})^{2}-w(D_{AB}^{j+1,j})^{\dagger
}(1-b_{0})^{2}$ and $\Pi =i(v+\frac{\gamma }{2})D_{AB}^{jj}b_{0}+i(v-\frac{%
	\gamma }{2})(D_{AB}^{jj})^{\dagger
}b_{0}-iwD_{AB}^{j+1,j}(1-b_{0})-iw(D_{AB}^{j+1,j})^{\dagger }(1-b_{0})$. It can be diagonalized using a combined squeezing and displacement transformation
\begin{widetext}
\begin{equation}
		\begin{split}
			H^{\mathrm{D}}& =e^{S^{\mathrm{d}}[\Lambda,\Pi]}e^{S^{\mathrm{sq}}[\Lambda]}H^{A,A^{2}}e^{-S^{\mathrm{d}}[\Lambda,\Pi]}e^{-S^{\mathrm{sq}}[\Lambda]}  \\
			S^\mathrm{d}[\Lambda,\Pi]& =\frac {g_{coupling}} {\sqrt L\omega_0}\left(\frac{\mathcal{W}[\Lambda]}{\omega_0}\right)^{-\frac32}\left(a^\dagger-a\right)\Pi,  \\
			S^\mathrm{sq}[\Lambda]& =\frac14\ln\left(\frac{\mathcal{W}[\Lambda]}{\omega_0}\right)\left(a^2-(a^\dagger)^2\right). 
		\end{split}
\end{equation}%
\end{widetext}
where $\mathcal{W}[\Lambda]=\omega _{c}\sqrt{1-2\frac{g_{coupling}^{2}}{L\omega _{c}}\Lambda }$. 
Thus, the
cavity ground state $\left\vert \phi ^{GS}\right\rangle $ is given by the
ground state of Hamitonian $H^{D}$ which is a displaced squeezed vacuum state $%
\left\vert \phi ^{GS}\right\rangle ^{D}$ that is connected to the bare
cavity vacuum $\left\vert 0\right\rangle $ through a combined squeezing and displacement transformation,

\begin{equation}
	\left\vert \phi ^{GS}\right\rangle ^{D}=e^{S^{\mathrm{d}}[\Lambda,\Pi]}e^{S^{\mathrm{sq}}[\Lambda]}\left\vert 0\right\rangle ,  \label{17}
\end{equation}%
Therefore, we come to the fact that the cavity ground state is a displaced squeezed vacuum state in the thermodynamic limit when $\mathcal{PT}$ symmetry is not broken.

\bibliography{apssampNotes}

\end{document}